\documentclass[aps, prb, amsmath, amssymb, reprint,floats,floatfix]{revtex4-2}

\usepackage{xcolor}

\usepackage{graphicx}

\usepackage[export]{adjustbox}

\usepackage[colorlinks]{hyperref}

\usepackage{bm}

\usepackage{braket}

\usepackage{enumerate}
\usepackage[caption=false]{subfig}
\usepackage{comment}
\usepackage{physics}
\begin{document}
\author{Chun Y. Leung}
\address{Department of Physics, Lancaster University, Lancaster LA1 4YB, United Kingdom}
\email{c.y.leung2@lancaster.ac.uk}

\author{Alessandro Romito}
\address{Department of Physics, Lancaster University, Lancaster LA1 4YB, United Kingdom}
\email{alessandro.romito@lancaster.ac.uk}
\date{\today}

\title{Entanglement and operator correlation signatures of many-body quantum Zeno phases in inefficiently monitored noisy systems}
\begin{abstract}
     The interplay between information-scrambling Hamiltonians and local continuous measurements hosts platforms for exotic measurement-induced phase transition in out-of-equilibrium steady states. Here, we consider such transitions under the addition of local random white noise and measurement inefficiency in a XX spin chain. We identify a non-monotonic dependence on the local noise strength in both the averaged entanglement and operator correlations, specifically the subsystem parity variance. While the non-monotonicity persists at any finite efficiency for the operator correlations, it disappears at finite inefficiency for the entanglement. 
     The analysis of scaling with the system size in a finite length chain indicates that, at finite efficiency, this effect leads to distinct MiPTs for operator correlations and entanglement. Our result hints at a difference between area-law entanglement scaling and Zeno-localized phases for inefficient monitoring.
\end{abstract}
\maketitle

\section{Introduction}

The ability to control quantum dynamics using quantum measurement has driven monitored systems to the centre of active research as a promising platform to host novel phases of matter far from equilibrium. 
At their heart, these systems feature a competition between an observer actively performing measurements on the system and coherent unitary dynamics~\cite{jacobs2014quantum,wiseman2009quantum}. 
The simplest implementation of this competition is the Zeno effect in which sufficiently strong monitoring freezes the unitary dynamics, locking the system in an eigenstate of the measured observable~\cite{misra1977zeno,peres1980zeno,PhysRevResearch.2.033512}.
In many-body systems, where unitary evolution generically scrambles information and has been shown to induce unbounded volume law scaling of entanglement, the competition with local monitoring leads to novel phenomena, most remarkably measurement-induced phase transitions (MiPTs) in the entanglement scaling ~\cite{10.21468/SciPostPhys.7.2.024,PhysRevX.9.031009,PhysRevB.100.134306,PhysRevB.99.224307,PhysRevB.100.064204,fisher2023random}.

Random quantum circuits offer a general platform for exploring generic entanglement MiPTs in which the measurement-inherent stochasticity naturally combines with the randomness of unitary gates. 
In these settings, dynamics is independent of the specifics of the local entangling component. 
Random quantum circuits punctuated by measurements have been intensively studied and shown to host MiPT from volume to area law entanglement~\cite{fisher2023random}, with some experimental evidence~\cite{noel2022measurement,koh2023measurement,google2023measurement}.
MiPTs with continuous Hamiltonian unitary dynamics have been studied prevalently for deterministic Hamiltonians~\cite{10.21468/SciPostPhys.7.2.024,alberton2021entanglement,10.21468/SciPostPhys.14.3.031,PhysRevX.13.041046,PhysRevB.103.224210,PhysRevB.102.054302,PhysRevB.110.035126,piccitto2024impact,PhysRevB.108.104313,PhysRevB.109.144306,PhysRevB.109.L060302,gal2023entanglement,paviglianiti2023enhanced,PhysRevB.102.035119,PhysRevA.102.033316,PhysRevResearch.2.013022,PhysRevB.104.184422,Zhang2022universal,PhysRevB.106.134206,PhysRevX.11.041004}, also in the presence of static (quenched) disorder~\cite{PhysRevB.108.165126,PhysRevB.110.024303,PhysRevB.107.L220201,chahine2023entanglement,PhysRevB.107.174203}. 
Stochastic unitary components, however, have been shown to modify the MiPTs, altering the scaling of entanglement~\cite{PhysRevX.13.041045,leung2023theory} or the critical point~\cite{PhysRevB.105.144202,PhysRevB.108.165126}. 
The stochastic unitary component and the measurement-induced one can also be regarded as different unravellings of the same averaged dynamics. 
From this perspective, averaging over the stochastic unitary or introducing a finite measurement inefficiency introduces a Lindblad dissipative term, washing away measurement-induced phase transitions~\cite{PhysRevResearch.4.033001,10.21468/SciPostPhys.12.1.009,carisch2024does,PhysRevResearch.5.L042031}. 

In this paper, we address the interplay of a stochastic unitary component and inherent measurement-induced stochasticity in stochastic Lindblad dynamics, from purity-preserving monitored systems to fully averaged deterministic Lindblad dynamics.
We study specifically the dynamics of a qubit chain with nearest-neighbour interactions and additional incompatible stochastic contributions from (i)  local continuous quantum measurement and (ii) random local unitary. 
We first address the dynamics in the simplest case of two qubits, revealing a non-trivial, non-monotonic behaviour in the entanglement and operator correlations dependence on the local unitary noise. 
With the introduction of inefficiency, this non-monotonicity in the entanglement disappears below a threshold efficiency value. On the contrary, it persists in the operator correlations  (for any finite efficiency), indicating a breakdown in the correspondence between entanglement scaling and a quantum Zeno phase signalled by correlations. 
We explore the implication of this breakdown for MiPTs by extending the protocol to a finite-length chain. 
The system size dependence of both entanglement and operator correlations indicates that the correspondence between the two, valid for fully efficient measurement, is broken with the inclusion of inefficiency. 
This breakdown suggests a difference between the measurement-induced quantum Zeno phase and area-law entanglement phase, with different phase diagrams obtained from entanglement and operator correlations.

The rest of the paper is structured as follows. In Sec.~\ref{section:model}, we present the model of interest. In Sec.~\ref{section:two qub}, we introduce various entanglement and operator correlation quantities and discuss and analyse numerically the simplest version of the model (a 2-qubit system). We demonstrate here how operator correlations in this system are insensitive to measurement inefficiency contrary to the behaviour shown in entanglement. In Sec.~\ref{section:chain}, we extend our analysis to a spin- $1/2 \mathrm{XX}$ chain, demonstrating that operator correlations and entanglement can lead to different measurement-induced phase transitions. We summarise our results and possible implications of our work in Sec.~\ref{section:discussion}. 

\section{Model}\label{section:model}
\begin{figure}
    \centering
    \includegraphics[width=0.45\textwidth]{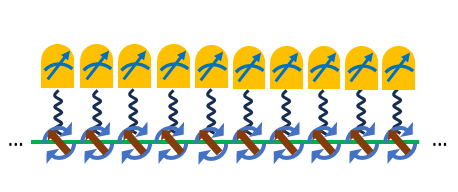}
    \caption{Sketch of the model under consideration. A spin $1/2$ chain (red arrows) with nearest neighbour spin-flip is subjected to local continuous measurement of $\sigma^z$ (yellow detectors). The spins are subject to random local magnetic fields in $y$-direction (blue arrow). The two stochastic dynamics are incompatible.}
    \label{fig:model both 2 qubits and chain}
\end{figure}

We study an XY spin-$1/2$ chain of length $L$ subject to local continuous measurements of the $z$-component of the spin and under the influence of a local random traverse magnetic field in the $ y$ direction. 
A sketch of the model is presented in fig.\ref{fig:model both 2 qubits and chain}.
We model the local random magnetic field as a local white noise statistically independent at different sites and the measurement backaction via quantum state diffusion equations in which we explicitly incorporate measurement inefficiency. The overall dynamics of the system can then be written as\cite{jacobs2014quantum}
\begin{align}\label{eq:mixed dynamics}
    &d\rho_t= -idt[H,\rho_t] \\
    & -i\sum_{j=1,L}[\sigma_j^y,\rho_t]d\xi_j^t -\frac{\Gamma}{2}dt\sum_{j=1,L}\left[\sigma_j^y,[\sigma_j^y,\rho_t]\right] \nonumber \\
    &+\sum_{j=1,L}\{\sigma_z-\langle\sigma_j^z\rangle,\rho_t\}\sqrt{\eta}dW_j^t-\frac{\lambda}{2}dt\sum_{j=1,L}\left[\sigma_j^z,[\sigma_j^z,\rho_t]\right] \nonumber,
\end{align}
where 
\begin{equation}
    H=\sum_j i\sigma_j^{+}\sigma_{j+1}^{-}+\, h.c..
\end{equation}
We set the notation so that we denote $\sigma_j^{\alpha} \, , \alpha\in\{x,y,z,+,-\}$ as the $\alpha$ Pauli operator on site $j$.  The state's labelling follows the usual convention with $0$ for spin-down states and $1$ for spin-up states so that e.g. $\ket{01}$ represents a state with spin-down on the first site and spin-up on the second site.
Furthermore, In Eq.~\eqref{eq:mixed dynamics}, $\eta \in [0,1]$ quantifies the efficiency of the measurements, $\Gamma$ is the strength of the white noise and $\lambda$ is the measurement strength. $d\xi_j^t$ and $dW_j^t$ are independent Itô processes, with $d\xi_j^td\xi_{j'}^{t'}=\Gamma dt\delta_{t,t'}\delta_{j,j'}$ and $dW_j^tdW_{j'}^{t'}=\lambda dt\delta_{t,t'}\delta_{j,j'}$. 
The strength of the Hamiltonian can be fixed as it merely appears as an overall energy scale that we set to be $1$ hereafter. 

The efficiency of the quantum diffusion process is controlled by $\eta$, which vanishes for completely inefficient measurements when  Eq.~\eqref{eq:mixed dynamics} reduces to a Lindbladian master equation for the measurement part.
As described Appendix~\ref{appendix:inefficient}, the inclusion of inefficiency in the quantum diffusion equation can be obtained from a microscopic model in which each site is coupled to an ancillary degree of freedom, which in turn is projectively measured \cite{jacobs2014quantum,PhysRevB.103.224210,rouchon2022tutorial,wiseman2009quantum}. 
Inefficiency is then included as the lack of some fraction of the measurement readouts --- a common uncontrolled error in experiments \cite{jacobs2014quantum,wiseman2009quantum,rouchon2022tutorial}. 
This necessarily induces mixedness in the density matrix, with the associated complications in the quantification of entanglement.

It is important to note that detecting non-trivial Zeno regimes or capturing the entanglement dynamics requires computing averages of non-linear observables over different quantum trajectories. 
Averages over measurement outcomes (i.e. over quantum trajectories) and noise realisations are denoted by an overline above $\overline{\dots}$.
If one considers linear observables e.g. $\overline{\Tr[\sigma_j^z\rho_t]}=\Tr[\sigma_j^z\overline{\rho}_t]$ or observable of mean state $\Tr[\sigma^z_j\overline{\rho_t}]$, the dynamics is entirely controlled by $\overline{\rho}_t$, which is determined by a Lindbladian
\begin{align}\label{eq:mean density matrix}
    &d\overline{\rho}_t=-idt[H,\overline{\rho}_t] -\frac{\Gamma}{2}dt\sum_{j=1,2}\left[\sigma_j^y,[\sigma_j^y,\overline{\rho}_t]\right] \nonumber \\ &-\frac{\lambda}{2}dt\sum_{j=1,2}\left[\sigma_j^z,[\sigma_j^z,\overline{\rho}_t]\right].
\end{align}
In this case, the long-time steady state $\overline{\rho}_{t\to\infty}\sim \mathbb{I}$ is the fully mixed state. 
Instead, averages of non-linear observables e.g. $\overline{\Tr[\sigma_j^z\rho_t]^2}$ contain non-trivial statistical correlation terms leading to non-trivial steady state value; this is analogous to deep thermalisation which is only detected by the higher moment of density matrix along each quantum trajectory\cite{ippoliti2022solvable}.

\section{Two qubits}\label{section:two qub}

To elucidate our motivation and results, we begin by presenting the most simple scenario of the model: a 2-qubit system (cf Eq.\ref{eq:mixed dynamics} with $j\in \{1,2\}$).
We use the model to introduce quantifiers of entanglements and operator correlations, as well as proxies for them, which will be used later for the extensive system. 
We are particularly interested in the case of inefficient measurements in which the state is generically non-pure, and entanglement quantifiers for pure states, like entanglement entropy, are no longer applicable.

\subsection{Entanglement and operator correlationsmeasures}\label{subsection:Ent and corr stuff}

\paragraph{Concurrence}
There are several proposed estimators of entanglement in an overall mixed state; for two qubits, a natural choice is the Concurrence $\mathcal{C}$ which is a genuine entanglement monotone and remains valid for mixed state\cite{hill1997entanglement}. 
It is defined as follow: let $\rho_t$ be the instantaneous 2-qubits density matrix at time $t$, we define $\Tilde{\rho}_t=\sigma^y\otimes\sigma^y\rho_t^*\sigma^y\otimes\sigma^y$ and a non-Hermitian matrix $\rho_t\Tilde{\rho}_t$. The Concurrence $\mathcal{C}$ associated is
\begin{align}\label{eq:Concurrence}
    \mathcal{C}=\max\left(0,\sqrt{\lambda_1}-\sqrt{\lambda_2}-\sqrt{\lambda_3}-\sqrt{\lambda_4}\right), 
\end{align}
where $\lambda_1 \dots \lambda_4$ are the the eigenvalues of the matrix $\rho_t\Tilde{\rho}_t$ in descending order. $\mathcal{C}=0$ corresponds to no entanglement e.g. product states, while $\mathcal{C}=1$ represents maximal entanglement e.g. Bell pairs.

\paragraph{Negativity}
As $\mathcal{C}$ only applies to a 2-qubit system, other entanglement monotones should be considered for later extension to a chain. A good candidate is the subsystem logarithmic negativity (an entanglement monotone \cite{plenio2005logarithmic}), which can readily be applied to larger systems. The subsystem logarithmic negativity is defined as:
\begin{align}\label{eq: negativity}
    \epsilon_{A}=\log||\rho^{T_A}||,
\end{align}
where $\rho^{T_A}$ denotes the partial transposition of the density matrix $\rho$ concerning region $A$ (in this case, one of the two qubits), and $||\rho^{T_A}||=\Tr[\sqrt{\rho^{T_A \,\dagger}\rho^{T_A}}]$ is the sum of the singular value of $\rho^{T_A}$.

\paragraph{subsystem parity variance}
Operator correlations are quantities which are closely related to entanglement~\cite{zeng2019quantum}. For example, in many-body physics, gapped area law entanglement phases are associated with exponentially decaying 2-point correlations, whilst logarithmic growth entanglement is associated with power-law decaying 2-point correlations. However, they capture both classical and quantum correlations in the system. We are interested in operator correlations that signal a quantum Zeno regime in which the system is frozen in an eigenstate of the measured observable.

There are several candidates to be considered.
Here, we choose the subsystem parity variance, which quantifies how close a state is to a polarised spin up/down state \cite{PhysRevResearch.4.033001,bao2021symmetry,PhysRevB.108.214302}. 
 It is defined as
\begin{align}\label{eq:half system parity}
    P_{1/2}=\langle \prod_{j=1}^{L/2}\sigma_j^z\rangle^2,
\end{align} 
and, for a two qubits system, it is merely
\begin{align}\label{eq: sub parity}
    P_{1}=\Tr[\sigma_1^z\rho_t]^2.
\end{align}
It is clear how this measure serves as an indicator for the quantum Zeno effect: under frequent measurements (spin-$z$ in our model), spin excitations are localised, becoming closer to a product state of spin-up/spin-down states.
Therefore, a high subsystem parity variance indicates a quantum Zeno regime.

\paragraph{subsystem purity}
For completeness, we also compute the half-system purity. 
The half systems purity is defined as
\begin{align}\label{eq: sub purity}
    \mu_{1/2}=\Tr[\rho_{1/2}^2],
\end{align}
where $\rho_{1/2}$ is the reduced density matrix of one part of the system.

Physically, $\mu_{1/2}$ is not an entanglement monotone, including classical and quantum correlations. 
Its relation to the quantum Zeno effect is clear: if spin excitation is localised, the half-system reduced density matrix is highly pure and does not correlate much with the rest of the system.

\subsection{Results --- efficient measurements}\label{section:2q efficient results}


We compute the entanglement monotones and operator correlation functions introduced in Sec.~\ref{subsection:Ent and corr stuff} for the two-site chain by numerical simulation of Eq.~\eqref{eq:mixed dynamics}, following the procedure in Ref.~\onlinecite{PhysRevA.91.012118}. We set $\delta t=\mathrm{min}(0.05,0.05/\lambda,0.05/\Gamma)$ across all simulations, which guarantees that the continuous limit is reached (tests with smaller time steps leave the results unaffected). 
For numerical convenience, we also restrict the initial state of the form 
\begin{equation}\label{eq:state}\ket{\psi}=\alpha\ket{00}+\beta\ket{01}+\gamma\ket{10}+\delta\ket{11},
\end{equation}
with $\alpha,\beta,\gamma,\delta \in \mathbb{R}$ so that they remain real at all times according to the evolution in Eq.\eqref{eq:mixed dynamics}.

Before proceeding to the results, we shall discuss briefly some of the effects of the various contributions. In a 2-qubit system, $H$ with $j\in\{1,2\}$ is the usual hopping term coupling the two qubits. In the absence of any randomness, starting from an initial state $\ket{\psi}=\alpha\ket{00}+\beta\ket{01}+\gamma\ket{10}+\delta\ket{11}$, the system displays periodicity in entanglement reflecting the unitarity of $H$. 
With the addition of white noise and measurement, which does not commute with $H$, all three dynamics compete. 
Without measurement, finite local white noises scramble information within the system, suggesting a noise's strength-independent entanglement in the long-time steady-state dynamics. With the addition of measurement (which tends to localise information), entanglement is expected to be suppressed as the measurement strength increases. The ultimate fate of entanglement and correlations with the interplay of all three dynamics depends non-trivially on their relative strength.

\begin{figure*}
    \centering
    \includegraphics[width=1\textwidth]{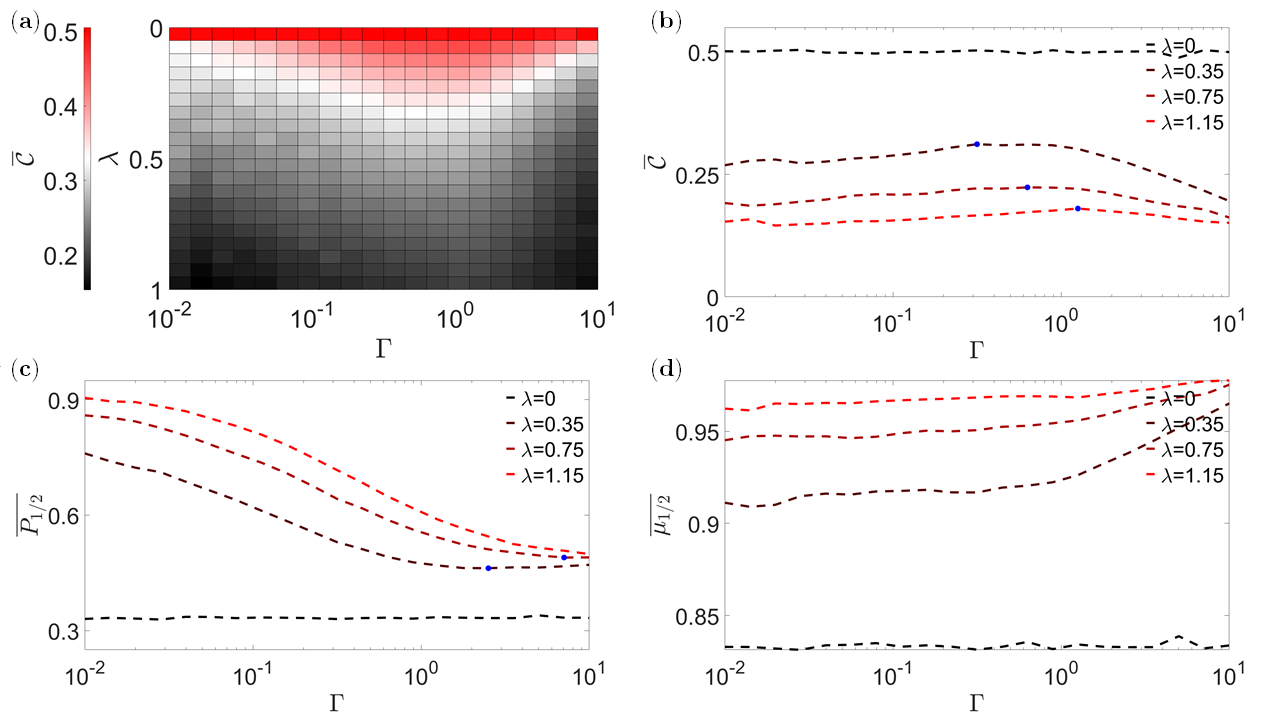}
    \caption{Average entanglement and subsystem parity variance in the steady state of the 2 qubits model (cf Eq.\eqref{eq:mixed dynamics}). (a): density plot of the average concurrence $\overline{\mathcal{C}}$ over an array of noise strengths $\Gamma$ and measurement rates $\lambda$. A non-monotonic dependence along $\Gamma$ can be observed. (b): horizontal cuts along the density plot fig.\ref{fig:2q pure}(a) displaying the average $\overline{\mathcal{C}}$ as a function of $\Gamma$ for various $\lambda$, see legend. The blue dots indicate the maximum of each curve. (c): average half system parity $\overline{P_{1/2}}$ as a function of $\Gamma$ for various $\lambda$. The blue dots indicate the minimum of each curve, and for $\lambda=1.15$ (red) the minimum lies outside of the plot (estimated to be $\Gamma\approx20$). (d): average half system parity $\overline{P_1}$ as function of $\Gamma$ for various $\lambda$. The curves are monotonic.}
    \label{fig:2q pure}
\end{figure*}

First, we present our main results for efficient measurement (pure state dynamics) in Fig.~\ref{fig:2q pure}. 
Panels (a) and (b) display the results of the average concurrence $\overline{\mathcal{C}}$ in the long-time steady-state as a function of the noise strength $\Gamma$, for various measurement strength $\lambda$.

Without measurement ($\lambda=0$), the average concurrence in the long-time steady-state $\overline{\mathcal{C}}$ converges to a value independent of the noise strength, $\overline{\mathcal{C}}=0.5$. This is a direct consequence of the information scrambling by the local random unitary, which, in the steady state, leads to a flat probability distribution over all the allowed states. 
As a result, the noise strength merely affects how fast the information is scrambled (time required to saturate). At the same time, the steady-state value is uniquely determined by the subspace of available states. 
With the parametrization in Eq.~\eqref{eq:state}, the average concurrence is given by 
\begin{equation}
\overline{\mathcal{C}} =\int_\Omega dS P \vert \alpha \delta -\beta \gamma \vert=0.5
\end{equation}
where $\Omega$ is the hyper-surface defined by $\alpha^2+\beta^2+\gamma^2+\delta^2=1$, $P=1/(2\pi^2)$ is the normalised constant probability distribution and $dS$ the infinitesimal surface element.
With the inclusion of measurement, entanglement is overall suppressed displaying a trend of reduction with increasing measurement strength $\lambda$, as indicated in Fig.\ref{fig:2q pure}(a) (vertical slices) and (b). 
A non-monotonic behaviour in $\overline{\mathcal{C}}$ with increasing $\Gamma$, as a non-trivial result of the interplay between noise and measurement. 
The initial increase of average concurrence with $\Gamma$ for weak noise can be understood heuristically as an information scrambling effect from the random local unitary. 
This scrambling competes with and reduces the localising effect from measurement. 
This simple argument, however, breaks down when the noise is increased further: $\overline{\mathcal{C}}$ first reaches a maximum as indicated by the blue dots, then decreases for larger $\Gamma$. 
This is one of our first findings: competing local noise and measurement reduce entanglement for strong noise, contrary to enhancement for weak noise. 
The reduction in entanglement induced by measurement for strong noise can be understood as an effect of fast fluctuations of local energy levels, which hinder the ability of $ H_0$ to entangle adjacent spins.

This non-monotonic behaviour is observed in the subsystem parity variance as well. In fig.~\ref{fig:2q pure}(c), we observe from the half-system parity that there is an initial decrease for small $\Gamma$, reaching a minimum (blue dots), followed by an increase for larger $\Gamma$. 
The overall values of $\overline{P_{1/2}}$ in the presence of measurement are higher than the noise-only scenario, revealing less correlations within the system and the dynamics resembling closer to a quantum Zeno regime. The non-monotonic behaviour in $\overline{P_{1/2}}$ indicates various degrees of localised correlations, and it is qualitatively in agreement with the behaviour observed in the concurrence: high $\overline{\mathcal{C}}\leftrightarrow$ low $\overline{P_1}$ and vice versa. It is worth pointing out that the location of the minimum in $\overline{P_{1/2}}$ does not match exactly the location of the maximum in $\overline{\mathcal{C}}$.
In both cases, the maximum/minimum of the non-monotonicity shifts to a larger value of $\Gamma$ with increasing $\lambda$ as indicated by the blue dots. In particular, the minimum in $\overline{P_{1/2}}$ shifts faster than the maximum of $\overline{\mathcal{C}}$ (the minimum for $\lambda=1.15$ lies outside of the plot in fig.~\ref{fig:2q pure}(c), estimated to be $\Gamma\approx20$ ).

Interestingly, in fig.~\ref{fig:2q pure}(d), $\overline{\mu_{1/2}}$ does not show any non-monotonic behaviour in the set of $\lambda$'s values presented here, but it is present for smaller $\lambda$  as reported in  app.~\ref{appendix:more numerics}. 
Although the overall increase with $\Gamma$ qualitatively agrees with the overall trend observed in $\overline{\mathcal{C}}$ and $\overline{P_{1/2}}$, the disappearance of non-monotonicity for larger $\lambda$ suggests that different indicators, may have quantitative differences in capturing the features of Zeno dynamics.  In the following,  we will drop $\overline{\mu_{1/2}}$ and retain the subsystem parity variance, $P{1/2}$ that more closely matches the entanglement dynamics.

\subsection{Inefficient measurements}\label{subsection:2q ineffi}

As presented above, for efficient measurements (pure state), both the entanglement and correlations capture, to some extent, the same non-monotonic feature in the dynamics, indicating some correspondence between correlations and entanglement in a similar fashion to equilibrium physics. 
In a many-body setting, this suggests that a quantum Zeno regime, in which the dynamics stabilize a short-range correlated dark-state (measurement operator eigenstate), is related to low entanglement area law in the system \cite{PhysRevB.98.205136,Biella2021manybodyquantumzeno}. 
However, as demonstrated below, this relationship appears somewhat broken in inefficient measurements (mixed state).

\begin{figure*}
    \centering
    \includegraphics[width=1\textwidth]{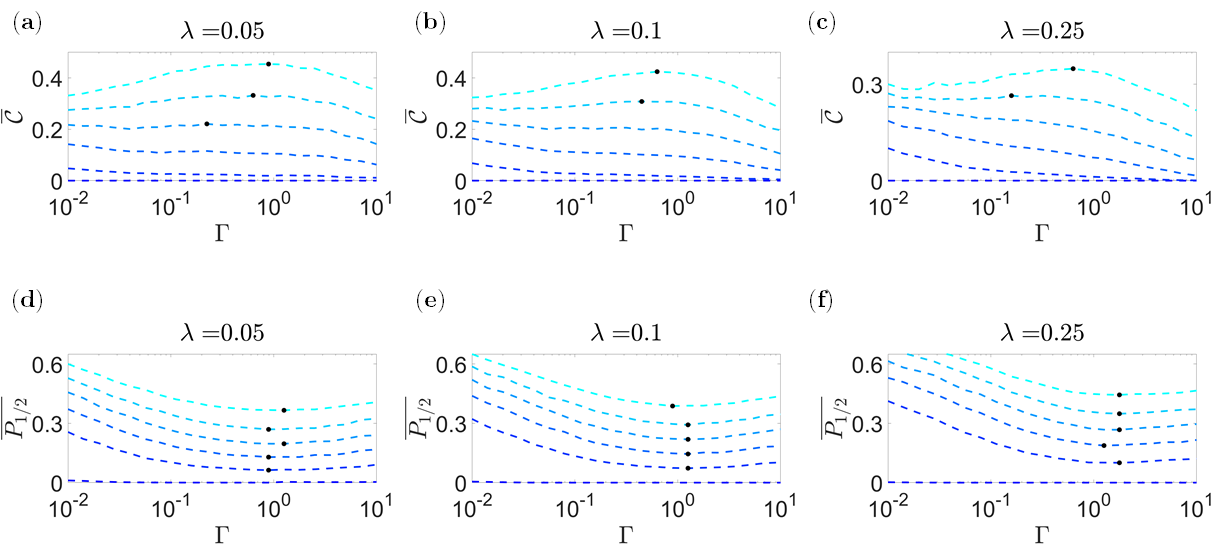}
    \caption{Trajectories averaged concurrence $\overline{\mathcal{C}}$ (top) and half-system parity $\overline{P_{1/2}}$ (bottom) as a function of local white noise strength $\Gamma$, for given measurement strength $\lambda$ with increasing values from left to right panels, and various measurement efficiencies $\eta$. The values of $\eta$, in decreasing shade, are $\{0,0.2,0.4,0.6,0.8,1\}$. (a)-(c): the black dots indicate the maximum for each curve. The maximum drifts to lower $\Gamma$ for smaller $\eta$. (d)-(f): the black dots indicate the minimum for each curve. The position of the black dots is essentially independent of $\eta$.}
    \label{fig:2q Concurrence ineffi}
\end{figure*}

In Fig.\ref{fig:2q Concurrence ineffi}, we display the results of the concurrence and the half-system parity variance for inefficient measurements. From the simulations of $\overline{\mathcal{C}}$ (fig.~\ref{fig:2q Concurrence ineffi}(a)-(c)), we observe that the entanglement in the system generally decreases with decreasing efficiency of the measurements (lighter to darker blue). 
This is a direct consequence of inefficient measurements that make the density matrix increasingly mixed and closer to the fully mixed state as the inefficiency increases, diminishing entanglement in the system. 
An important feature is observed here: for any measurement strength, $\lambda$, the non-monotonicity in $\Gamma$ disappears below a threshold efficiency $\eta^*$, which depends on $\lambda$ (disappearance of maxima indicated by black dots). 
We interpret this as a new regime in which the density matrix is highly mixed, and the scrambling from local random unitaries cannot out-compete information loss/localisation from local measurements.

However, the operator correlations in the system tell a different story. In Fig.\ref{fig:2q Concurrence ineffi}(d)-(f), we display the results of the average subsystem parity variance in the steady state as a function of $\Gamma$. Although the absolute values of $\overline{P_1}$ are lower for decreasing $\eta$, non-monotonic behaviour is present across all finite efficiency $\eta$.
This behaviour is different from that of the entanglement: the average concurrence $\overline{\mathcal{C}}$ becomes monotonically decreasing for larger $\eta$, whilst the average subsystem parity variance $\overline{P_{1}}$ remains non-monotonic. This comparison shows that entanglement and operator correlations may behave as two distinct system features. 
Therefore, it is natural to ask whether the phase transition captured by operator correlations is the same as that captured by entanglement. This will be the main theme of the next section, Sec.\ref{section:chain}

\section{Spin chain}\label{section:chain}

We now extend our investigation to a chain of more than two qubits (cf eq.\eqref{eq:mixed dynamics}). 
Given the qualitative differences of inefficiencies in entanglement and operator correlations highlighted in the last section, we focus specifically on whether the phase transitions (of a many-body system) indicated by the two separate measures are equivalent. 
We employ the half system logarithmic entanglement negativity labelled by $\epsilon_{L/2}\equiv\epsilon_{1/2}$ to quantify entanglement (cf eq.\eqref{eq: negativity}), and half system parity variance labelled by $P_{1/2}=\Tr[\prod_{j=1}^{L/2}\sigma_j^z]^2$ for operator correlations in the spin chain dynamics (cf Eq.\eqref{eq: sub parity}).

Note that although there exists a generalised many-body concurrence \cite{PhysRevA.64.042315,bhaskara2017generalized}, it only applies to pure state (efficient measurement), hence we employ the entanglement negativity as a proper entanglement estimator.

\subsection{Efficient measurement}

From Sec.\ref{section:2q efficient results}, entanglement and operator correlations generally agree with each other in capturing the same qualitative features for the case of two qubits for efficient measurement. 
We expect this to hold here as well.
\begin{figure*}
    \centering
    \includegraphics[width=1\textwidth]{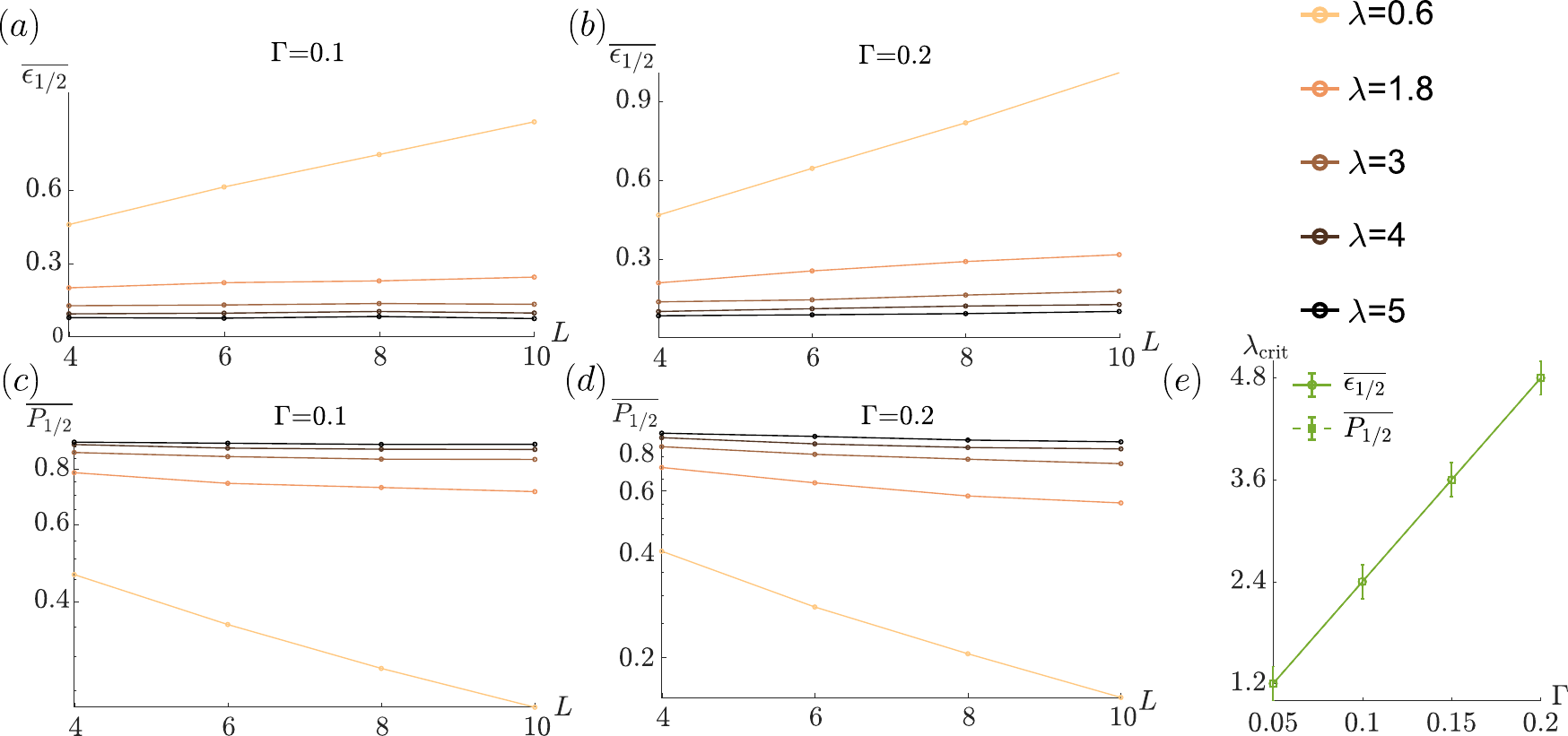}
    \caption[width=2\textwidth]{Scaling of the averaged entanglement negativity $\overline{\epsilon_{1/2}}$ (panels a,b) and half system parity (panels c,d) in the spin chain model under efficient continuous measurement $\eta=1$ for different values of the measurement strength. 
    The results are shown for two different values of noise strength, $\Gamma=0.1$ in panels (a) and (d), and $\Gamma=0.2$ in panels (b) and (c). (e): estimated critical measurement strength $\lambda_{\mathrm{crit}}$ as a function of $\Gamma$ as estimated from from $\overline{P_{1/2}}$ (square marker/dashed line) and $\overline{\epsilon_{1/2}}$ (triangle marker/solid line). The two lines are indistinguishable since they fully overlap}
    \label{fig:Chain efficient}
\end{figure*}
The system's entanglement changes from extensive-entanglement scaling (pale yellow lines) to area-law behaviour (dark copper lines) as $\lambda$ increases. Similarly, $\overline{P_{1/2}}$ changes from system-size dependent to system-size independent, indicating strong spin-spin correlations.

In Fig.~\ref{fig:Chain efficient}, we display the scaling of $\overline{\epsilon_{1/2}}$ and $\overline{P_{1/2}}$ with respect to different system sizes $L$, for various $\lambda$ at fixed $\Gamma$. 
Fig.\ref{fig:Chain efficient}(a) shows a qualitative change in entanglement scaling upon increasing measurement strength for a fixed value of $\Gamma=0.1$.
For small measurement strength $\lambda\leq 2.4$, $\overline{\epsilon_{1/2}}$ is $L$ dependent with extensive scaling of entanglement; in contrast, $\overline{\epsilon_{1/2}}$ becomes $L$ independent for larger $\lambda>2.4$, suggesting an area-law phase. 
When the noise strength is increased, the extensive entanglement phase sets on at increased values of $\lambda >4$ as shown in fig.~\ref{fig:Chain efficient}(b). 
Although the system sizes are limited and finite size effects are relevant, the results indicate noise strength-dependent MiPTs between an area-law phase and an extensive entanglement scaling phase. 
With the caveat of finite-size scaling, the latter appears to be a volume-law scaling phase. 
We also note that our results imply a measurement-induced phase transition induced by local unitary noise, which has recently been addressed in a different model~\cite{eissler2024unraveling}. 


Turning our attention to the results of half system parity in fig.~\ref{fig:Chain efficient}(c) and (b), their $L$-scaling is qualitatively consistent with that of entanglement: whenever $\overline{\epsilon_{1/2}}$ indicates an area law phase entanglement, $\overline{P_{1/2}}$ is $L$ independent (pale colour lines in fig.\ref{fig:Chain efficient}(c)), whereas it decreases with larger $L$ in the extensive entanglement phase (dark colour lines fig.\ref{fig:Chain efficient}(c) and all lines in fig.\ref{fig:Chain efficient}(b))\cite{PhysRevResearch.4.033001,bao2021symmetry}.

We can identify a critical measurement strength, which separates the two distinct phases from either the entanglement negativity or the subsystem parity variance. In the former, $\overline{\epsilon_{1/2}}$, it separates the extensive entanglement phase from the area-law phase and in the latter, $\overline{P_{1/2}}$, it separates long-range operator correlations from short-range operator correlations. 
We denote the respective critical measurement strengths by $\lambda_{c,\overline{\epsilon}}$ and $\lambda_{c,\overline{P}}$. 
Repeating the analysis in Fig.~\ref{fig:Chain efficient}(a-d) for different values of $\Gamma$, we can estimate the $\Gamma$-dependence of $\lambda_{c,\overline{\epsilon}}$ and $\lambda_{c,\overline{P}}$. 
Shown in Fig.\ref{fig:Chain efficient}(e),
$\lambda_{c,\overline{\epsilon}}$ and $\lambda_{c,\overline{P}}$ approximately coincide with each other and increase monotonically for increasing $\Gamma$.

\subsection{Inefficient measurements}

We now discuss our results for inefficient measurements. As observed from the simple case of two qubits (cf ec.~\ref{subsection:2q ineffi}), correlations and entanglement under inefficient measurements may display different behaviours. Therefore, we are interested in the implications of this discrepancy on the measurement-induced entanglement transition, comparing it with the transition indicated by correlations.

\begin{figure*}
    \centering
    \includegraphics[width=1\textwidth]{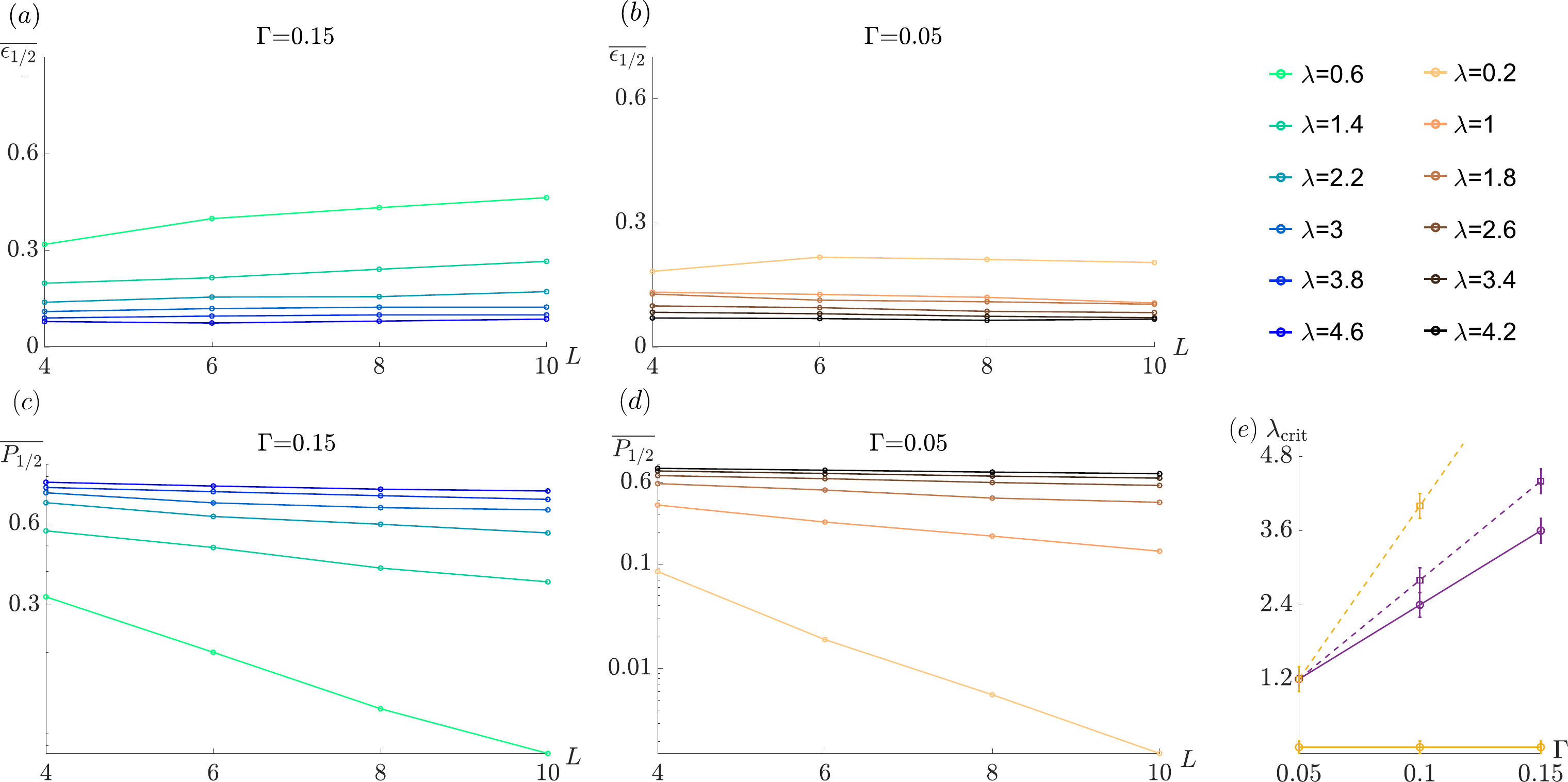}
    \caption{Scaling of the averaged entanglement negativity $\overline{\epsilon_{1/2}}$ (panels a,b) and half system parity (panels c,d) in the spin chain model under inefficient continuous measurement $\eta<1$. The results are shown for two different values of noise strength and inefficiencies, $\Gamma=0.15$ $\eta=0.8$ in panels (a) and (c), and $\Gamma=0.05$ $\eta=0.4$ in panels (b) and (d). Legend of the lines appears at the top left corner. (e): estimated critical measurement strength $\lambda_{\mathrm{crit}}$ as a function of $\Gamma$ for $\lambda_{c,\overline{P}}$ (square marker/dashed line) and $\lambda_{c,\overline{\epsilon}}$ (triangle marker/solid line). The colour scheme indicates different values of inefficiency with purple for$eta=0.8$ and yellow for $\eta=0.4$.}
    \label{fig:Chain inefficient}
\end{figure*}

In fig.~\ref{fig:Chain inefficient}(a)-(d), we report the results for the scaling of the average half-system negativity $\overline{\epsilon_{1/2}}$ and the average half-system parity variance $\overline{P_{1/2}}$ for inefficient measurements at given noise strength and inefficiency, for different measurement strengths.  For high efficiency, $\eta=0.8$, we observe that the behaviours of entanglement and operator correlations remain qualitatively similar to the fully efficient case, i.e. both $\overline{P_{1/2}}$ and $\overline{\epsilon_{1/2}}$ undergo a transition from long-range to short-range upon increasing the measurement strength. As an example, this is shown in fig.~\ref{fig:Chain inefficient} panel (a) and (c) for $\Gamma=0.15$. 
However, different from the fully monitored case, the estimated critical measurement strength from
the two indicators is different, with $\lambda_{c,\overline{\epsilon}}\approx 3.6 <\lambda_{c,\overline{P}} \approx 4.4$.
The critical points follow a dependence on the noise strength similar to that observed in the fully monitored case, with the critical measurement strength increasing with the noise strength, as shown in panel (e).
Notably the difference 
between $\lambda_{c,\overline{\epsilon}}$ and $\lambda_{c,\overline{P}}$ is also reduced at smaller noise strength, and the two are no longer distinguishable at $\Gamma \approx 0.05$.

The different behaviour of the entanglement negativity and the half-system parity variance mirrors the behaviours observed in two qubits, in which for $\eta>\eta^*$, the non-monotonicity survives in entanglement, but not in the subsystem parity variance
The discrepancy between the two is further enhanced as we reduce the efficiency to $\eta=0.4$ [panels (b) and (d)].  
In fact, $\lambda_c(\epsilon)$ appears at a much lower value and is no longer $\lambda$ dependent. This aligns with the known effect that highly inefficient measurement tends to thermalise the system with vanishing entanglement. This suggests a different kind of MiPT controlled by inefficiencies, which generates mixedness and suppresses entanglement differently from the localising effect of measurement~\cite{yu2009sudden}. Such transition to vanishing entanglement due to inefficiency was also found in other models, and in some cases, a critical inefficiency can be identified~\cite{PhysRevResearch.4.033001,10.21468/SciPostPhys.12.1.009}. 
Below the critical efficiency, the system is generally in the mixed phase. This transition to a mixed phase can also be observed through the effect of noise. For small inefficiency, increasing $\Gamma$ still has the effect of favouring the extensive entanglement phase, which increases $\lambda_{c,\overline{\epsilon}}$ as shown in fig.~\ref{fig:Chain inefficient} (e) (solid purple line). However, as the inefficiency is increased further, this entanglement enhancing effect by noise is suppressed as displayed in fig.~\ref{fig:Chain inefficient} (e) (solid yellow line), where $\lambda_{c,\overline{\epsilon}}$ is $\Gamma$-independent and possibly equal to 0 for any finite measurement strength. 
The absence of this enhancement by noise is in line with the discussion previously on entanglement suppression by inefficiency/thermalisation in Sec.~\ref{section:two qub}.


Finally, we note that the case $\eta=0$ is trivial: any finite $\lambda$ will induce trivial dynamics since the measurement part of the master equation Eq.\eqref{eq:mixed dynamics} reduces down to a Lindbladian, and the density matrix at long times is merely proportional to identity.

\section{discussion and conclusions}\label{section:discussion}

In this work, we have numerically investigated the effect of local unitary noise in locally monitored systems. 
In a minimal 2-qubit setup, the interplay of these competing dynamics produces an intriguing non-monotonic behaviour in entanglement and operator correlations as a function of noise strength. 
This unique feature associated with quantum trajectory dynamics is most visible at small measurement strength, where the system displays higher entanglement/correlations for intermediate noise strength. 
With increasing measurement strength, the minimum/maximum of the non-monotonicity shifts to larger noise strength and the non-monotonicity becomes less prominent. 

Interestingly, upon the inclusion of measurement inefficiency, the 2-qubit system signals non-trivial dynamics for the entanglement and operator correlations: entanglement gradually becomes monotonic, whereas correlations, specifically the hal-parity variance, remain non-monotonic for all finite measurement inefficiencies. This suggests a breakdown of the conventional correspondence between entanglement and operator correlations present in equilibrium physics, and mixed dynamics could host correlations, which behave differently from entanglement. 
This can be heuristically understood because operator correlators generally capture both quantum and classical correlations, which might be responsible for the breakdown. Such behaviour was also hinted at in pure state unitary dynamics~\cite{foligno2023temporal}, where entanglement entropy scales differently from higher order Rényi entropy. 

Motivated by this observation in mixed state dynamics (inefficient measurement), we study an extended spin $1/2$ chain model and compute the system size scaling of entanglement and half-parity variance. 
For efficient measurements, the scaling of entanglement and operator correlations display the typical correspondence in behaviour so that the short-range (area-law) entanglement phase coincides with the short-range phase from operator correlations. For inefficient measurements, however, the breakdown of the relation is visible: short-range entanglement no longer corresponds to short-range correlations. 
The results suggest that entanglement and operator correlations can generically differ from each other in mixed-state dynamics, and the phase diagrams indicated by the two quantities are no longer equivalent.

Our results hint at a richer scenario for MiPTs in mixed states than that depicted by the entanglement phase transition alone. 
They further raise the question of how generic the reported discrepancy between the entanglement scaling and operator correlations is.
On the one hand, the generality of the observed features beyond the model studied here is an interesting aspect to address, especially for other models where a mixed state transition has been identified~\cite{PhysRevResearch.4.033001}.
On the other hand, the question extends to whether the discrepancy can be somehow associated with some classical correlation components in the subsystem parity variance and whether non-classical correlations different from entanglement, like quantum discord, may play a role.

\section{acknowledgement}\label{section:ack}
The authors would like to thank G. Kells for useful discussions. This work is supported by
the Royal Society, grant no. IECR2212041.

\appendix

\section{Inefficiency}\label{appendix:inefficient}

This appendix outlines the derivation of the quantum state diffusion in the presence of inefficiency Eq.~\eqref{eq:mixed dynamics}.
From a microscopic perspective, there are two main ways in which measurement is considered inefficient: 1. the detector only records a fraction of the readout, and 2. the readout by the detector has a non-zero probability of being wrong. Case 1 arises naturally in a photon counting detector, while case 2 is a fault in the experimental apparatus\cite{jacobs2014quantum}. 

Consider case 1, where one can view the real physical detector as two imaginary sub-detectors: the first sub-detector with measurement streng th $\lambda_{(1)}$ is a perfect detector where all readouts from the real detector are recorded by it solely (cf. Eq.~\eqref{eq:mixed dynamics} for perfect efficiency). 
In contrast, the second sub-detector with measurement strength $\lambda_{(2)}$ loses all its readouts, and the observer only has access to the average dynamics induced by it (c.f. Eq.~\eqref{eq:mean density matrix}). Representing the measurement operator as $O_j$, one can then write down the combined effects of these two sub-detectors on the conditional density matrix as from \eqref{eq:mixed dynamics}  and Eq.~\eqref{eq:mean density matrix}:
\begin{align}\label{eq: ineff app rho}
    d\rho_t&=\{O_j-\langle O_j\rangle , \rho_t\}dW_{(1)}^t - \frac{\lambda_{(1)}}{2}dt\left[O_j,[O_j,\rho_t]\right] \nonumber \\ & - \frac{\lambda_{(2)}}{2}dt\left[O_j,[O_j,\rho_t]\right],
\end{align}
where $dW_{(1)}^tdW_{(1)}^{t'}=\lambda_{(1)}dt\delta_{t,t'}$. Calling a new Itô process $dW^tdW^{t'}=(\lambda_{(1)}+\lambda_{(2)})dt\delta_{t,t'}$, Eq.\eqref{eq: ineff app rho} precisely describes a real detector with measurement strength $\lambda=\lambda_{(1)}+\lambda_{(2)}$ and efficiency $\eta=\lambda_{(1)}/(\lambda_{(1)}+\lambda_{(2)})$.

To demonstrate case 2, we remind ourselves that continuous measurements on the spin are simply feedback from the projective measurements performed on the ancillae via a quantum channel upon tracing out the ancilla degree of freedom. Restricting to a single two-level ancilla, the channel may be represented by the following Kraus' operators:
\begin{align}\label{eq: ineff app bino}
    K_u&=\frac{1}{\sqrt{2}}\left(\sqrt{1+\epsilon}\ketbra{1}{1}+\sqrt{1-\epsilon}\ketbra{0}{0}\right) \nonumber \\
    K_d&=\frac{1}{\sqrt{2}}\left(\sqrt{1-\epsilon}\ketbra{1}{1}+\sqrt{1+\epsilon}\ketbra{0}{0}\right),
\end{align}
and they satisfy the condition $K_u^{\dagger}K_u+K_d^{\dagger}K_d=\mathbb{I}$. $\epsilon$ is a small parameter, and we need only terms up to $\mathcal{O}(\epsilon^2)$. The two Kraus' operators in Eq.\eqref{eq: ineff app bino} correspond to the two possible readouts from the ancilla and they update the state in the following way: given a readout $\mathrm{r}\in\{u,d\}$, the state after a measurement event is
\begin{align}\label{eq: ineff app perfect}
    \rho_{(u)}&=\frac{K_u\rho K_u^{\dagger}}{p_u} \ , \text{if $\mathrm{r}=u$} \nonumber \\
    \rho_{(d)}&=\frac{K_d\rho K_d^{\dagger}}{p_d} \ , \text{if $\mathrm{r}=d$},
\end{align}
with respective probability $p_u=\Tr[K_u\rho K_u^{\dagger}]$ and $p_d=\Tr[K_d\rho K_d^{\dagger}]$. $\rho_{u}$ and $\rho_{d}$ represents the post-measurement state corresponding to readout $u$ and $d$. Eq.\eqref{eq: ineff app perfect} assumes perfect detector: let $\Delta\in(0,1)$ be the conditional probability of a detector readout to be $u$ given that the real readout is $u$, $\Delta=p(\mathrm{r}=u|u)=1$ is unity for perfect measurement (respectively for $d$). However, 
Consider now systematic errors so that $\Delta=p(\mathrm{r}=u|u)<1$ can be less than 1. Translating the effect of this error onto the density matrix update, Eq.\eqref{eq: ineff app perfect} is modified to be:
\begin{align}\label{eq: ineff app wrong}
    \rho_{(\mathrm{r}=u|u)}&=\frac{ \Delta K_u\rho K_u^{\dagger} + (1-\Delta)K_d\rho K_d^{\dagger}}{ p(u) } \nonumber \\
    \rho_{(\mathrm{r}=d|d)}&=\frac{\Delta K_d\rho K_d^{\dagger} + (1-\Delta)K_u\rho K_u^{\dagger}}{ p(d) }.
\end{align}
$p_{(u)}=\Tr[\Delta K_u\rho K_u^{\dagger} + (1-\Delta)K_d\rho K_d^{\dagger}]$ (and similarly $p_{(d)}$) is now modified. $\Delta=0.5$ corresponds to a completely inefficient detector whose readout is completely random, and $\Delta<0.5$ is equivalent to exchanging $u\leftrightarrow d$ with a new $\Delta'=1-\Delta$ (a detector with $\Delta=0$ has its readout `flipped', but working just fine). Expanding Eq.\eqref{eq: ineff app wrong} up to $\mathcal{O}(\epsilon^2)$, we can combine both equations into a single differential equation:
\begin{align}\label{eq: ineff app raw}
    d\rho&=\frac{dW}{2}(2\Delta-1)\{\sigma^z,\rho\}-\frac{\epsilon^2}{8}\left[ \sigma^z,[\sigma^z,\rho]\right] \nonumber \\&-dW(2\Delta-1)\langle \sigma^z\rangle\rho-\frac{(2\Delta-1)^2}{2}\epsilon^2\langle\sigma^z\rangle\{\sigma^z,\rho\}\nonumber \\&+(2\Delta-1)^2\epsilon^2\rho\langle\sigma^z\rangle^2.
\end{align}
$d\rho$ represents the change in $\rho$ after a measurement event ($d\rho\equiv\rho_{(\mathrm{r}=k|k)}-\rho, k\in\{u,d\}$) and we introduce a binomial variable $dW=\pm\epsilon$ with probability distribution $p(\pm\epsilon)=1/2(1\pm(2\Delta-1)\epsilon\langle\sigma^z\rangle)$ and $\overline{dW}=\epsilon^2(2\Delta-1)\langle\sigma^z\rangle$. Constructing a new binomial variable $\overline{d\xi}=dW-\overline{dW}$ (overline corresponds to average), the mean is now centred at zero $\overline{d\xi}=0$ and the variance $\overline{d\xi^2}=\epsilon^2 +\mathcal{O}(\epsilon^4)\equiv\overline{dW^2}$ is unchanged up $\mathcal{O}(\epsilon^2)$. Eq.\eqref{eq: ineff app raw} now becomes:
\begin{align}
    d\rho=\frac{d\xi}{2}(2\Delta-1)\{\sigma^z-\langle\sigma^z\rangle,\rho\}-\frac{\epsilon^2}{8}\left[\sigma^z,[\sigma^z,\rho]\right]
\end{align}
Setting the scaling $\epsilon=2\sqrt{\lambda dt}$, $d\xi$ is equivalent to a Wiener process (central limit theorem giving Gaussian distribution) with mean $0$ and variance $4\lambda dt$ in the time continuum limit. With this, we recover Eq.\eqref{eq:mixed dynamics} with $\eta=|2\Delta-1|^2$ (and appropriate scaling).

\section{Supplementary numerical simulations}\label{appendix:more numerics}
We present more numerical results for entanglement negativity and purity for completeness. We also report simulations using the quantum jump equation, which shows identical average features.

\paragraph{Quantum jump}---The procedure to simulate the quantum jump equation is slightly different from the one used in the main text: we modify the measurement operator $\sigma^z$ to a jump operator $\hat{n}_j=1/2(1+\sigma^z_j)$ which is a projector. The quantum jump equation can be derived using suitable Kraus' operators, similar to Eq.\eqref{eq: ineff app bino}:
\begin{align}\label{eq: QJ app Kraus}
    K_u&=\sqrt{\epsilon}\ketbra{1}{1} \nonumber \\
    K_d&=\sqrt{1-\epsilon}\ketbra{1}{1}+\ketbra{0}{0}.
\end{align}
Here $\epsilon$ is a small number quantifying the strength of the measurement, and its temporal scaling should be set as $\epsilon\sim dt\equiv\gamma dt$ to derive the time continuum quantum jump equation \cite{jacobs2014quantum,turkeshi2021measurement}. Measurement inefficiency is incorporated as outlined in App.~\ref{appendix:inefficient}: given a true readout $u$, the probability of the detector output being $u$ is not unity (cf. Eq\eqref{eq: ineff app perfect}).

\begin{figure}
    \centering
    \includegraphics[width=0.45\textwidth]{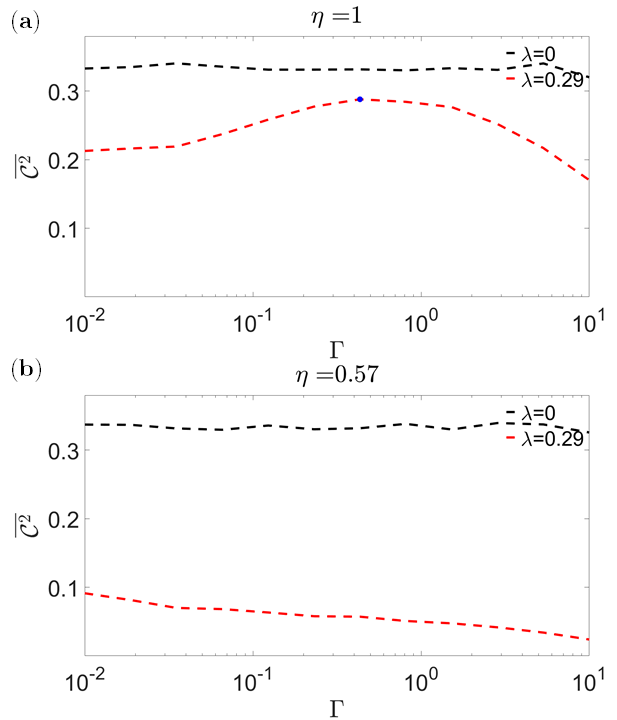}
    \caption{Average squared concurrence $\overline{\mathcal{C}^2}$ as a function of noise strength and for different values of the measurement strength for jump operator $\hat{n}_j=1/2(1+\sigma^z_j)$, $j\in\{1,2\}$. Results for two measurement inefficiencies are presented: (a) $\eta=1$ and (b) $\eta=0.57$.}
    \label{fig:2q QJ concur}
\end{figure}

In fig.~\ref{fig:2q QJ concur}, we present the results for average concurrence squared $\overline{\mathcal{C}^2}$ using the quantum jump equation. Non-monotonicity is present for perfect measurement (fig.~\ref{fig:2q QJ concur}(a)), and it disappears for sufficiently inefficient measurement (fig.~\ref{fig:2q QJ concur}(a)).

\begin{figure}
    \centering
    \includegraphics[width=0.45\textwidth]{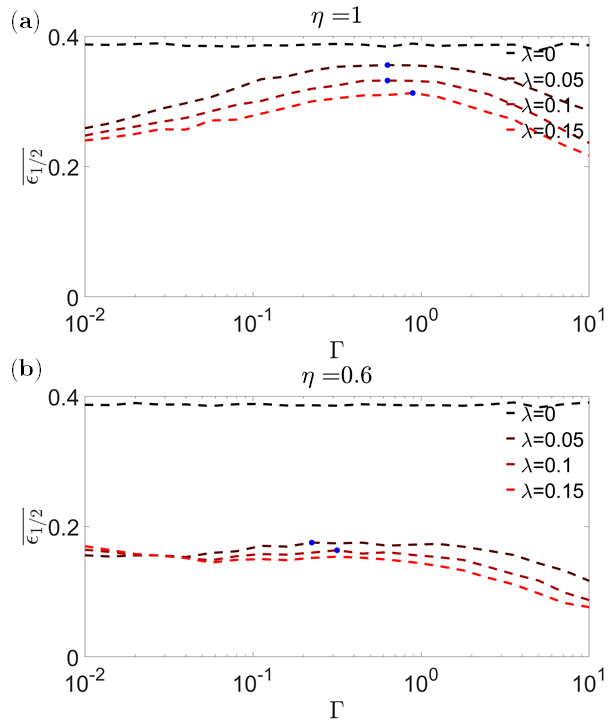}
    \caption{Average half system logarithmic negativity $\overline{\epsilon_{1/2}}$ as a function of noise strength for different values of the measurement strength, obtained by using identical time evolution as in the main text (cf Eq.\eqref{eq:mixed dynamics}). Results for two measurement inefficiencies are presented: (a) $\eta=1$ and (b) $\eta=0.6$. The presence (absence) of a blue dot indicates whether a non-monotonic behaviour is present(absence), and its location corresponds to the maximum.}
    \label{fig:2q negat}
\end{figure}

\paragraph{Logarithmic negativity}---In Fig.~\ref{fig:2q negat}, we display the results of half system logarithmic negativity, obtained by simulating Eq.\eqref{eq:mixed dynamics}. Non-monotonicity is also present, which confirms that this is a general entanglement feature in this 2-qubit system, irrespective of the monotone used for entanglement.

\pagebreak

\end{document}